\documentclass{article}
\usepackage[
  paperheight=8.5in,
  paperwidth=5.5in,
  left=10mm,
  right=10mm,
  top=20mm,
  bottom=20mm]{geometry}
\usepackage[utf8]{inputenc}
\usepackage{amsmath}
\usepackage{algorithm}
\usepackage{algpseudocode}
\usepackage{graphicx}
\usepackage{wrapfig}
\usepackage[bottom]{footmisc}
\usepackage{listings}
\usepackage{enumitem}
\usepackage{caption}
\usepackage{subcaption}

\usepackage{wrapfig}
\usepackage{ragged2e}

\usepackage{array}
\usepackage[table]{xcolor}
\usepackage{multirow}
\usepackage{booktabs}
\usepackage{hhline}
\definecolor{palegreen}{rgb}{0.6,0.98,0.6}

\usepackage{amsmath}
\usepackage{amssymb}
\usepackage{multicol}
\usepackage{lipsum}
\usepackage{hyphenat}
\PassOptionsToPackage{hyphens}{url}
\usepackage{url}

\usepackage{rotating}

\usepackage[T1]{fontenc}
\usepackage{textcomp}
\usepackage{listings}
\usepackage{setspace}

\newcommand*{\affmark}[1][*]{\textsuperscript{#1}}
\newcommand*{\email}[1]{\small{\texttt{#1}}}

\renewcommand{\footnoterule}{%
  \kern -3pt
  \hrule width \textwidth height 0.5pt
  \kern 2pt
}

\date{}

\usepackage{titlesec}
\titleformat*{\section}{\large\bfseries}
\titleformat*{\subsection}{\normalsize\bfseries}
\titleformat*{\subsubsection}{\normalsize\bfseries}

\usepackage{amsthm}
\newtheorem{theorem}{Theorem}[section]
\theoremstyle{definition}
\newtheorem{definition}{Definition}[section]
\usepackage[backend=bibtex,style=numeric]{biblatex}
\title{Demystifying the RSA Algorithm: An Intuitive Introduction for Novices in Cybersecurity\footnote{\protect Copyright \copyright 2022 by the Consortium for Computing Sciences in Colleges.
Permission to copy without fee all or part of this material is granted provided
that the copies are not made or distributed for direct commercial advantage,
the CCSC copyright notice and the title of the publication and its date appear,
and notice is given that copying is by permission of the Consortium for
Computing Sciences in Colleges.  To copy otherwise, or to republish, requires
a fee and/or specific permission.
}}

\author{
\affmark Zhengping Jay Luo\affmark [1], Ruowen Liu\affmark [2], \\Aarav Mehta\affmark [1] and Md Liakat Ali\affmark [1]\\
\affmark[1]Department of Computer Science and Physics\\
\affmark[2]Department of Mathematics\\
Rider University\\
Lawrenceville, NJ, 08648\\
\email{\{zluo,rliu,mdali\}@rider.edu}\\
}

\begin{document}
\maketitle

\begin{abstract}
Given the escalating importance of cybersecurity, it becomes increasingly beneficial for a diverse community to comprehend fundamental security mechanisms. Among these, the RSA algorithm stands out as a crucial component in public-key cryptosystems. However, understanding the RSA algorithm typically entails familiarity with number theory, modular arithmetic, and related concepts, which can often exceed the knowledge base of beginners entering the field of cybersecurity. In this study, we present an intuitively crafted, student-oriented introduction to the RSA algorithm. We assume that our readers possess only a basic background in mathematics and cybersecurity. Commencing with the three essential goals of public-key cryptosystems, we provide a step-by-step elucidation of how the RSA algorithm accomplishes these objectives. Additionally, we employ a toy example to further enhance practical understanding. Our assessment of student learning outcomes, conducted across two sections of the same course, reveals a discernible improvement in grades for the students.
\end{abstract}

\section{Introduction}
The three most widely accepted security goals of cybersecurity are shorted as ``CIA triad'', which stands for \textbf{C}onfidentiality, \textbf{I}ntegrity and \textbf{A}vailability. Cryptographic algorithms play a pivotal role in achieving confidentiality through private-key and public-key cryptographic algorithms. Public-key cryptographic algorithms, exemplified by the RSA algorithm, also contribute significantly to attaining another vital security goal—non-repudiation, particularly crucial in scenarios like electronic mail, where digital signatures are employed. Remarkably, the RSA algorithm was originally designed to address both confidentiality and non-repudiation goals in electronic mail \cite{rivest1978method,wahab2021hiding}.

Developed by Ron \textbf{R}ivest, Adi \textbf{S}hamir, and Leonard \textbf{A}dleman at the Massachusetts Institute of Technology (MIT) in 1976, the RSA algorithm stands as a pioneering implementation of the public-key cryptosystem, conceptualized by Diffie and Hellman \cite{diffie2022new}. Operating with two keys—a private key and a public key—the RSA algorithm facilitates secure communication. For instance, when two parties, Alice and Bob, aim to exchange messages covertly, Alice encrypts the message $M$ using Bob's public key, creating ciphertext $C$. This ciphertext is then sent to Bob, who decrypts it with their private key to retrieve the original plaintext $M$.

While this process may appear straightforward, generating the public and private keys involves intricate mathematical concepts such as number theory and modular arithmetic. These topics often pose challenges for beginners in cybersecurity, especially undergraduate students. In our work, we offer an intuitive and accessible perspective on understanding the RSA algorithm. Beginning with the three primary goals the RSA algorithm aims to achieve, we employ a student-oriented approach to elucidate the step-by-step design of the system. We acknowledge the potential lack of background knowledge in readers regarding number theory, modular arithmetic etc., and hence, we aim to simplify the mathematical rigor to make the content more approachable. 

Additionally, we provide a practical toy example of the RSA algorithm to enhance readers' understanding. Towards the end of the paper, we present a real-world student learning outcome assessment conducted on students from two different sections of the same course. Our results demonstrate that the proposed student-oriented approach outperforms the traditional method of explaining the RSA algorithm in terms of assignment grades.

The paper is organised as follows: the necessary foundational information of the RSA algorithm is provided in Section \ref{sec:background}. Then the detailed student-oriented style introduction of the algorithm is elaborated in Section \ref{sec:RSA}. In Section \ref{sec:example} we employed a specific toy example to demonstrate how to encrypt and decrypt the message in RSA from a practical perspective. We concluded the paper in Section \ref{sec:conclusion}.

\section{Background and Preliminaries}\label{sec:background}
In this section, we provide necessary background that gives the context and mathematical foundations of the RSA algorithm. Readers can also skip this section and use this section as a reference while reading Section \ref{sec:RSA}.

\begin{figure*}[h]
\centering
\begin{subfigure}[t]{.5\textwidth}
  \centering
  \includegraphics[scale = 0.23]{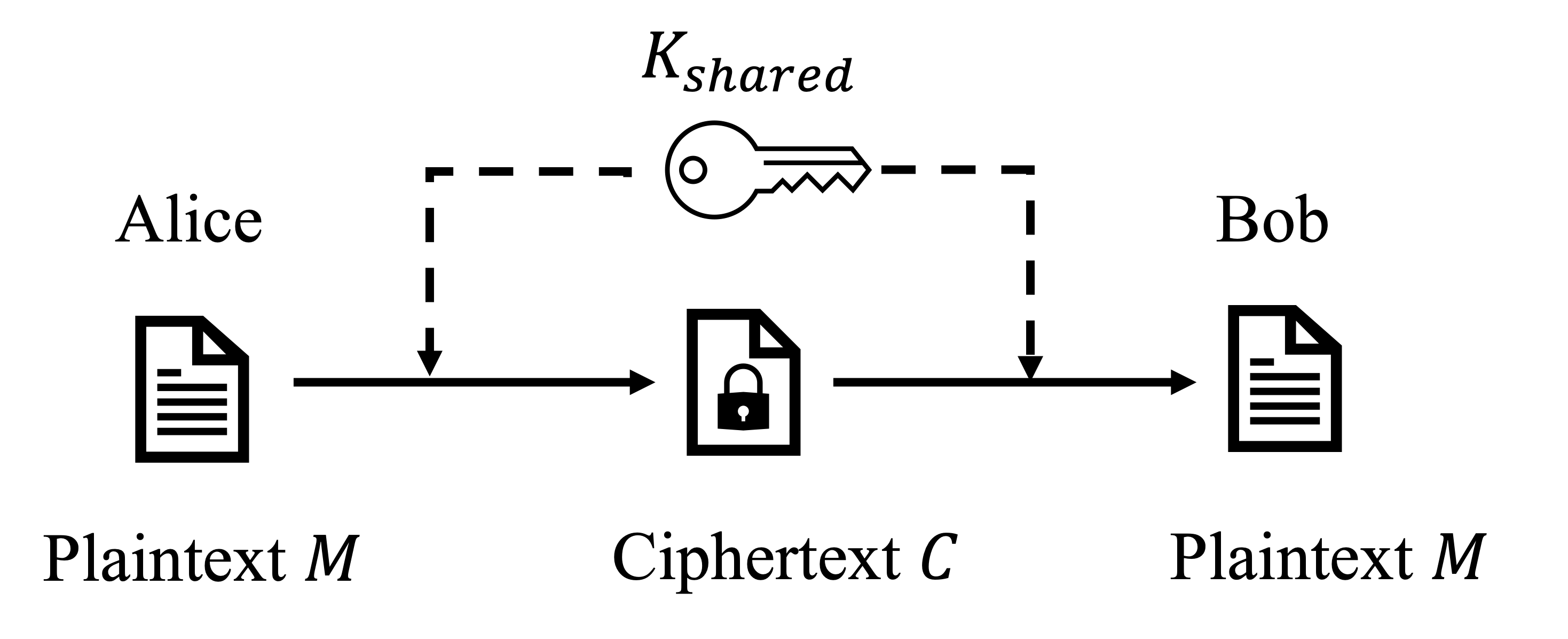}
  \caption{Symmetric-key cryptography}
\end{subfigure}\hfill 
\begin{subfigure}[t]{.5\textwidth}
  \centering
  \includegraphics[scale = 0.23]{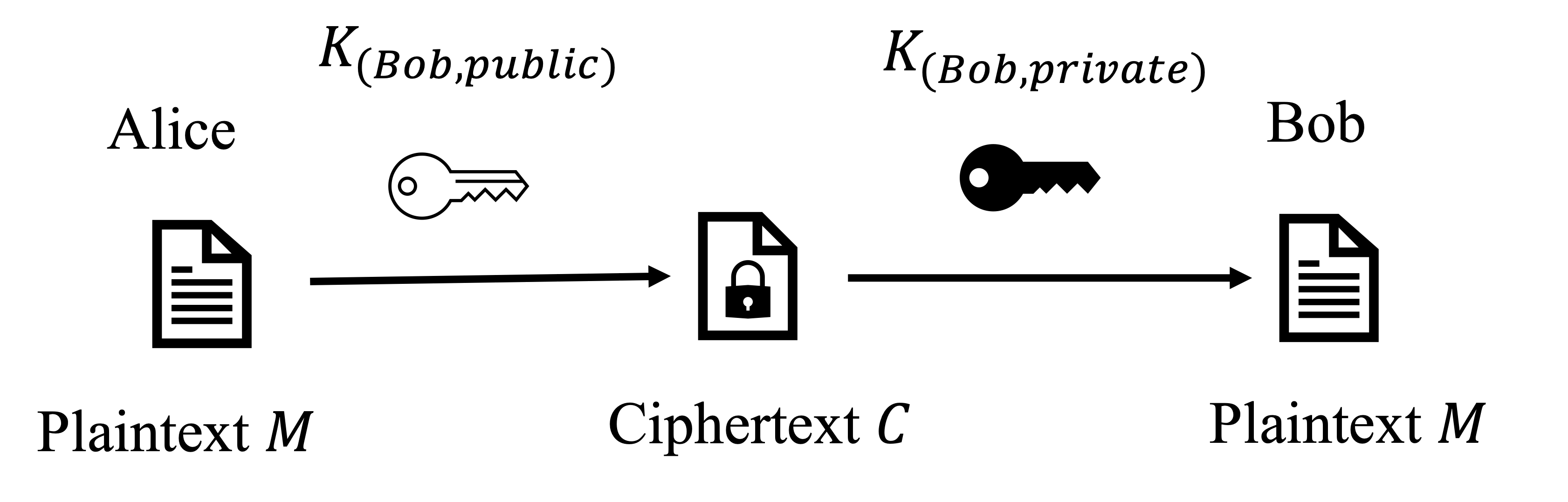}
  \caption{Public-key cryptography}
\end{subfigure}
\caption{The information flow when Alice sends a message to Bob using symmetric and public key cryptography.}
\label{fig:architecture}
\end{figure*}

\subsection{Symmetric-key and Public-key Cryptosystems}\label{sec:pkc}
One of the major challenges modern cryptographies want to address is how to ensure two end users, let's say Alice and Bob, could secretly exchange messages $M$ in an open and potentially unsafe environment. We have two strategies to tackle this challenge\cite{imam2021systematic}. 

The first strategy is to let both Alice and Bob share a secret key $K_{shared}$ and make sure any one of them can encrypt the plaintext $M$ into ciphertext $C$ using $K_{shared}$, while the other can recover $M$ from $C$ using the same key $K_{shared}$. This strategy is also known as symmetric-key cryptography \cite{anusha2020symmetric}. It is similar with a real-world padlock example in which we use a key to lock a cabinet. When someone wants to open the cabinet, they need to get the same key to unlock the padlock. The process of Alice using the symmetric-key cryptography to send a message to Bob is shown in Fig. \ref{fig:architecture}(a). 

One of the major problems with the symmetric-key cryptography is that end users have to share the same key in advance, which is often impractical in modern communication systems such as computer networks due to: : 
\begin{itemize}
\item In computer network systems, communication connections are usually random and instantaneously. Requiring a shared key among all the communication connections would be costly; 
\item Any information of the shared key sent over the open environment could be intercepted by malicious attackers, which will put the encryption out of work. Therefore, it is unrealistic to require all end users to share the same secret key in advance when they want to exchange information.  
\end{itemize}

In 1976, Diffie and Hellman \cite{diffie2022new} proposed the second strategy named as public-key cryptosystems to tackle these challenges. The basic idea is that both Alice and Bob will still share the same cryptograhic algorithm, but they no longer need to share the same secret key. Instead, the system will maintain two keys: a private key and a public key. The private key is only known to the owner while the public key can be accessed by anyone who wants to communicate with the owner. 

Every time if Alice wants to send a message to Bob, Alice will use Bob's public key $K_{(Bob, public)}$ to encrypt the message $M$. On Bob's side, the ciphertext $C$ can be decrypted using Bob's private key $K_{(Bob, private)}$. Since only Bob has $K_{(Bob, private)}$, thus no one else could recover $M$. The process of Alice using the public-key cryptosystem to send a message to Bob is shown in Fig. \ref{fig:architecture}(b). 

In this system, the two communication entities no longer need to communicate a shared key in advance, which addresses the major problem in symmetric-key cryptography. However, one of the major disadvantages is the public-key cryptography algorithms is usually more computationally costly than symmetric-key cryptography algorithms \cite{katz2020introduction,fotohi2020securing,liestyowati2020public}. 

The public-key cryptosystem is similar with our self-service drop box mechanism used in shipping industry. Anyone can put an envelope or a package (messages) into a public drop box (public key) provided by the shipping company (anyone could use the receiver’s public key to encrypt the message in public-key cryptosystems). However, only authorised personnel (receiver) from the shipping company that has the key (private) could open the drop box to get the mails/packages. 

Using public-key cryptosystems, two end users will no longer be required to share a secret key in advance when they need to exchange information. All the sender needs to know is the public key of the receiver and the cryptographic algorithm the receiver used, both of which are public information. The RSA algorithm is an implementation of the public-key cryptosystem concept. 

\subsection{Modular Arithmetic}\label{sec:modular}

Modular arithmetic is a branch of arithmetic for integers, where numbers ``wrap around'' when reaching a certain value. If we have a modulus $n$, which is an integer larger than 1, $a \hspace{0.1cm} \text{mod} \hspace{0.1cm}n $ is the remainder of $a$ divided by $n$.  For example, $7\hspace{0.1cm} \text{mod} \hspace{0.1cm} 3 = 1$. The result of $a \hspace{0.1cm} \text{mod} \hspace{0.1cm}n $ for any number $a$ will always be less than $n$ and greater than or equal to 0, i.e., $0 \leq a \hspace{0.1cm} \text{mod} \hspace{0.1cm} n <n$. In our $7\hspace{0.1cm} \text{mod} \hspace{0.1cm} 3 = 1$ example, obviously $1 < 3$.  If $a < n$, then $a \hspace{0.1cm} \text{mod} \hspace{0.1cm}n $ will always equal to $a$ itself. For example, $5\hspace{0.1cm} \text{mod} \hspace{0.1cm} 9 = 5$. In the case where integers $a$ and $b$ have the same remainder when divided by $n$, i.e., $a \hspace{0.1cm} \text{mod} \hspace{0.1cm}n = b \hspace{0.1cm} \text{mod} \hspace{0.1cm}n$, we have the following definition:

\begin{definition}
If $a$ and $b$ are integers and $m$ is a positive integer, then $a$ is \textit{congruent} to $b$ modulo $m$ if $m$ divides $a - b$. We use the notation $a \equiv b (\text{mod} \hspace{0.1cm} m)$ to indicate that $a$ is congruent to $b$ modulo $m$.
\end{definition}

For example, as 24 and 14 have the same remainder when divided by 5, we call 24 and 14 are congruent modulo 5, which can be represented as $24 \equiv 14 (\text{mod} \hspace{0.1cm} 5)$. In modular arithmetic, we use "$\equiv$" rather than "$=$" to denote the equivalence of modulo results. There is an important theorem of congruence that we will use in explaining the RSA algorithm:

\begin{theorem} 
If $a \equiv b (\text{mod} \hspace{0.1cm} m)$ for integers $a, b$ and $m$, then $ak \equiv bk (\text{mod} \hspace{0.1cm} m)$ and $a^k \equiv b^k (\text{mod} \hspace{0.1cm} m)$ for any integer $k$.
\end{theorem}

\begin{proof}
This can be proved by the definition of congruence. Since $a \equiv b (\text{mod} \hspace{0.1cm} m)$, then $a \hspace{0.1cm} \text{mod} \hspace{0.1cm} m =  b \hspace{0.1cm} \text{mod} \hspace{0.1cm} m$, i.e., $a - c_{1}m = b - c_{2}m$ for integers $c_{1}$ and $c_{2}$. Further this can be written as $a - b = cm$ for an integer $c$. We multiply both sides by an integer $k$ to get $ak - bk = ckm$, and perform modulo $m$ on both sides will get $ak \hspace{0.1cm} \text{mod} \hspace{0.1cm} m = bk \hspace{0.1cm} \text{mod} \hspace{0.1cm} m$, i.e., $ak \equiv bk (\text{mod} \hspace{0.1cm} m)$, which completes the proof. We can use similar strategies to prove  $a^k \equiv b^k (\text{mod} \hspace{0.1cm} m)$ for any integer $k$. 
\end{proof}
Another important theorem that we will use in proving the RSA algorithm is Bézout's theorem, 

\begin{theorem}[Bézout's theorem]
If $a$ and $b$ are positive integers, then there exist integers $s$ and $t$ such that the greatest common divisor of $a,b$, i.e., $gcd(a,b)$, can be represented as $gcd(a,b) = sa + tb$.
\end{theorem}

The detailed proof of this theorem can be found in \cite{rosen2011elementary}. The pair of $s$ and $t$ could be found using the \textit{Extended Euclidean Algorithm}. For example, $gcd(24,14) = 3 \times 24 + (-5) \times 14$. Now we give the definition of modular multiplicative inverse.

\begin{definition}
If there exist integers $a,b$ such that $ab \equiv 1 (\text{mod} \hspace{0.1cm} m)$, then $b$ is said to be an inverse of $a$ modulo $m$ and vice versa. 
\end{definition}

Based on this definition of modular multiplicative inverse and Bézout's theorem, we can derive the following theorem: 

\begin{theorem}
An inverse of $a$ modulo $m$ is guaranteed to be existed whenever $a$ and $m$ are relatively prime. 
\end{theorem}

\begin{proof}
As $a$ and $m$ are relatively prime, $gcd(a,m) = 1$. According to Bézout's theorem, there are integers $s$ and $t$ such that $gcd(a,m) = sa + tm = 1$. This implies that $sa + tm \equiv 1 (\text{mod} \hspace{0.1cm} m).$ As $tm \equiv 0 (\text{mod} \hspace{0.1cm} m),$ it follows that $sa \equiv 1 (\text{mod} \hspace{0.1cm} m). $ Consequently, $s$ is an inverse of $a$ modulo $m$. 
\end{proof}

To simplify the readability, we leave the proofs of these properties, such as the Extended Euclidean Algorithm in modular arithmetic, to the reader's interest. For those who wish to explore modular arithmetic and related theorems and proofs in greater depth, please refer to \cite{rosen2019discrete} for a detailed explanation.

\subsection{Prime Factorisation}\label{sec:factor}

Prime factorization means the decomposition, if possible, of a positive integer into a product of prime integers. For example, the prime factorization of 15 is $3 \times 5$, in which both 3 and 5 are prime numbers. Prime factorization is an important problem in number theory because still no efficient enough way has been discovered to find the prime factorization of an extremely large integer with existing classical computer systems. 

The RSA algorithm embeds prime factorization in its design to ensure there exists no efficient way to decipher the ciphertext in non-quantum computing systems. However, it does not mean that we would not find an efficient way to perform prime factorization in the future based on nowadays computer technology (a lot of mathematicians are still working on this problem); it also does not mean that we would not find an efficient way on future computers, such as quantum computing \cite{national2019quantum,hidary2019quantum,easttom2022quantum}.  In fact, an efficient way to perform prime factorization on quantum computers has already been found \cite{shor1994algorithms}. The problem is that a workable quantum computer is still estimated to be at least decades away \cite{bernstein2017post}. Therefore, we can safely say the RSA algorithm is secure at least for the time being.

\subsection{Euler's Theorem}\label{sec:euler}
Before introducing Euler's theorem, let's first provide the definition of Euler's totient function:

\begin{definition}
The Euler's totient function $\phi(.)$ is the number of positive integers that are less than and relatively prime to this integer, i.e., $\phi(n) = \text{the number of integers in} \{1,2,3, …, n-1\}  \text{which are relative prime to } n$.
\end{definition}

For example, given an integer 8, there exist four integers $1,3,5,7$ that are relatively prime to 8, thus Euler's totient function value $\phi(8) = 4$. You might have already realised that Euler's totient function value for a prime number $n$ is always $n - 1$, i.e., $\phi(n) = n - 1$, as all the $n-1$ positive integers less than $n$ are relative prime to $n$.  An important mathematical property of Euler's totient function is that:

\begin{theorem} 
If $m$ and $n$ are relatively prime integers, then $\phi(mn) = \phi(m) \times \phi(n)$.
\end{theorem}
For example, $\phi(6) = \phi(2) \times \phi(3) = 1 \times 2 = 2$. We’ll skip the proof here and the detailed proof of this theorem can be found in \cite{rosen2011elementary}. This property offers a convenient way to calculate Euler's totient function value if an integer can be factorized into the product of two prime numbers $m$ and $n$. In this case $\phi(mn)= \phi(m) \times \phi(n) = (m-1)(n-1)$ as $m, n$ are also relatively prime to each other, which we will use later in proving the RSA algorithm. The challenge here is that no efficient way has been found on modern computers to do prime factorization (as discussed in Section \ref{sec:factor}). 

It is worth noting that the complexity of prime factorization and computing the Euler’s totient function is equivalent for arbitrary integers. Essentially, both require evaluating whether the integer is relative prime to all the positive integers less than it. Therefore, it is also computationally difficult to calculate Euler’s totient function for large enough integers. Now we’re ready to introduce Euler's Theorem.

\begin{theorem}[Euler's Theorem]
If two integers $a$ and $n$ are relatively prime, i.e., $gcd(a,n) = 1$, and $n > 0$, then $a^{\phi(n)}\hspace{0.1cm} \equiv 1 \hspace{0.1cm}(\text{mod}\hspace{0.1cm} n)$.
\end{theorem}
For example, let $a=3$ and $n=4$, then they are relatively prime and we have $\phi(4) = 2$. Further we have $3^{\phi(4)} = 3^2 = 9$, thus, $3^{\phi(4)} \equiv 9 \equiv 1 \hspace{0.1cm}(\text{mod}\hspace{0.1cm} 4)$. We leave the proof of Euler's theorem to the readers due to the abundance of online resources on this topic \cite{rosen2011elementary}. It is worth noting that Euler's theorem provides a fast way to calculate $a^{\phi(n)}\hspace{0.1cm}\text{mod}\hspace{0.1cm} n$ when $a, n$ are relatively prime. This property plays a significant role in the RSA algorithm as we will see in the following section. 

After all the background information introduction, now we’re ready to start the introduction of the RSA algorithm, which is an implementation of the public-key cryptosystem.

\section{The RSA algorithm}\label{sec:RSA}
The RSA system was introduced in 1976. Now it is one of the most widely used public-key encryption methods in computer networks. To materialise a public-key cryptosystem, as we introduced in Section \ref{sec:pkc}, we want to achieve the following three basic goals \cite{rivest1978method}:
\begin{enumerate}
    \item\textbf{Efficiency:} The encryption and decryption process should be easy to compute for legitimate users who have the required key information.
    \item \textbf{Plaintext recovery:} We should be able to get the original plaintext $M$ through decrypting the ciphertext $C$.
    \item \textbf{Computational difficulty:} Without the private key information, there is no known efficient way to perform the decryption process.
\end{enumerate}

These three goals are critical in the success of the public-key systems. With these three goals in mind, we introduce the core encryption and decryption process of the RSA algorithm. The corresponding ciphertext $C$ of the plaintext $M$ is computed from 
\begin{equation}\label{encryption}
    C \equiv M^e \hspace{0.1cm}(\text{mod} \hspace{0.1cm} n).
\end{equation}
$e$ and $n$ is the public key information of the receiver. The decryption process is similar, which is
\begin{equation}\label{decryption}
    M' \equiv C^d \hspace{0.1cm}(\text{mod} \hspace{0.1cm} n).
\end{equation}
The private key information consists of $d$ and $n$. We use $M'$, not $M$ directly in Eq. (\ref{decryption}) because we want to highlight that this is the result we obtained from the decryption process. We will ensure $M' = M$ in the \textit{plaintext recovery} goal. 

Suppose Alice wants to send a secret message $M=2$ to Bob using the RSA algorithm. Bob's public key $(e,n)$ is $(113, 143)$ and the corresponding private key $(d,n)$ is $(17, 143)$, which means that the ciphertext $C \equiv M^e \hspace{0.1cm} (\text{mod} \hspace{0.1cm} n) \equiv 2^{113} \hspace{0.1cm} (\text{mod} \hspace{0.1cm} 143) \equiv 19$. Alice will send out $C = 19$ to Bob. Bob can then decrypt the ciphertext to recover the plaintext through $M^{'} \equiv C^d \hspace{0.1cm} (\text{mod} \hspace{0.1cm} n) \equiv 19^{17} \hspace{0.1cm} (\text{mod} \hspace{0.1cm} 143) \equiv 2$, which achieved the goal of $M^{'} = M$. The detailed encryption and decryption process of the RSA algorithm is shown as follows in Algorithm \ref{algorithm}.
\begin{algorithm*}
\caption{The encryption and decryption process of the RSA algorithm.}
\begin{algorithmic}[1]
\State \textbf{The Receiver}:
\State \quad \quad Choose two large random prime numbers $p$ and $q$ privately.
\State \quad \quad Obtain $n$ and $\phi(n)$ through $n = p \cdot q$ and $\phi(n) = (p-1)(q-1)$, then keep $p$ and $q$ in private or destroy them. 
\State \quad \quad Choose a large number $e$ that is relatively prime to $\phi(n)$. 
\State \quad \quad Compute $d$ such that $ed \equiv 1 (\text{mod} \hspace{0.1cm} \phi(n))$.
\State \quad \quad Release $(e,n)$ to the public and keep  $(d,n)$ as the private key.
\State \textbf{The Sender}:
\State \quad \quad Encrypt the message $M$ using the receiver's public key $(e,n)$, $C \equiv M^e \hspace{0.1cm}(\text{mod} \hspace{0.1cm} n)$, and send the ciphertext $C$ to the receiver. 
\State \textbf{The Receiver}:
\State \quad \quad Decrypt the received ciphertext $C$ using their own private key $(d,n)$ to recover $M' \equiv C^d \hspace{0.1cm}(\text{mod} \hspace{0.1cm} n)$.
\end{algorithmic}
\label{algorithm}
\end{algorithm*}

We now need to understand what conditions must be satisfied and how this process could achieve the three goals mentioned above. We will explain each goal with the associated conditions as follows.

\subsection{Goal 1: Efficiency}
Both encryption and decryption procedures are identical from an implementation perspective, making them straightforward to implement in practice. Additionally, private and public keys can be determined using standard and efficient methods on modern computers \cite{moriarty2016pkcs}.

We also need to be able to find $M^e \hspace{0.1cm}(\text{mod} \hspace{0.1cm} n)$ and $C^d \hspace{0.1cm}(\text{mod} \hspace{0.1cm} n)$ efficiently without using an excessive amount of memory given that $e,d,n$ are all large numbers. Directly computing the exponentiation operation of $M^e$ or $C^d$ is impractical, as their results can be very extremely large and require significant memory to store. Fortunately, this problem can be addressed using the fast modular exponentiation algorithm, which reduces the computational complexity to a logarithmic level. The detailed algorithm is provided in \cite{rosen2019discrete}.  

However, despite the RSA algorithm's careful design for efficiency, it is generally accepted that public-key cryptosystems are usually less efficient than symmetric-key cryptosystems. Therefore, in real-world scenarios, the RSA algorithm is primarily used for delivering the pre-shared key in symmetric-key cryptosystems, which is often a short message. When encrypting large amounts of information, symmetric-key cryptosystems are still preferred for their efficiency  \cite{katz2020introduction}.

\subsection{Goal 2: Plaintext Recovery}\label{sec:goal}

The second goal is to guarantee the accurate recovery of original plaintext $M$ from ciphertext $C$ using receiver's private key $(d,e)$, i.e., to ensure $M' = M$. Substituting $C$ in the encryption process as shown in Eq.(\ref{encryption}) to the decryption process as shown in Eq.(\ref{decryption}), it yields 
\begin{equation}\label{eq:recover}
    M' \equiv [M^e \hspace{0.1cm}(\text{mod} \hspace{0.1cm} n)]^d \equiv M^{ed} \hspace{0.1cm}(\text{mod} \hspace{0.1cm} n).
\end{equation}
As we know from Section \ref{sec:modular}, $M$ could also be written as
\begin{equation}
    M \equiv M \hspace{0.1cm} (\text{mod} \hspace{0.1cm} n), \hspace{0.1cm} \text{if} \hspace{0.1cm} M < n.
\end{equation}
Therefore, the goal can be reinterpreted as finding the conditions to guarantee
\begin{equation}\label{eq:goal}
M^{ed} \equiv M \hspace{0.1cm}(\text{mod} \hspace{0.1cm} n), \text{with}\hspace{0.1cm} M<n.
\end{equation}
As long as $M < n$, the above equation will hold. According to Euler's theorem (Section \ref{sec:euler}), if $M$ and $n$ are relatively prime, then $M^{\phi(n)} \equiv 1 \hspace{0.1cm}(\text{mod}\hspace{0.1cm} n)$.  By the modular arithmetic properties (Section \ref{sec:modular}), we can raise both sides to the $k$-th power, with $k$ being a positive integer, to get $M^{k\phi(n)} \equiv 1^k \equiv 1\hspace{0.1cm}(\text{mod}\hspace{0.1cm} n)$. Multiplying both sides by $M$ yields,
\begin{equation}\label{eq:m}
M^{k\phi(n)+1} \equiv M \hspace{0.1cm}(\text{mod}\hspace{0.1cm} n).
\end{equation}
Comparing Eq.\eqref{eq:recover} to Eq.\eqref{eq:m}, to ensure the correct recovery $M'=M$, we would now require 
\begin{equation}
M^{ed} \hspace{0.1cm}(\text{mod} \hspace{0.1cm} n) = M^{k\phi(n)+1}(\text{mod} \hspace{0.1cm} n) 
\end{equation}
i.e., we need 
\begin{equation}\label{eq:final}
    ed = k \phi(n)+1 \hspace{0.1cm}, k \hspace{0.1cm} \text{is a positive integer}.
\end{equation}

Up until now, we found that we have two conditions need to be satisfied in order to make above equations hold: (1) $M < n$ and (2) $M$ and $n$ are relatively prime. As long as these two conditions are satisfied, the above derivation from Eq.\eqref{eq:recover} to Eq.\eqref{eq:final} will hold. To satisfy the first condition, in real world, after choosing the large positive number $n$, we need to break long messages into small blocks such that each block can be represented as an integer $M$ that is less than $n$. We will explain how to ensure the second condition in Section \ref{sec:goal 3}. 

We now know that if we could find a pair of $e,d$ such that $ed = k\phi(n)+1$,$k$ is a positive integer. The two conditions for $M$ and $n$ are satisfied, then we’re confident that the original plaintext $M$ could be recovered from $C$. In the next section, we’ll see how these conditions are met and at the same time the \textit{computational difficulty} goal is also achieved. 

\subsection{Goal 3: Computational Difficulty}\label{sec:goal 3}
Now the challenge is reduced to a problem of finding appropriate values of $e$ and $d$, which are the major components of the public and private key respectively. The only clue we have now is $ed = k\phi(n)+1$, where $k$ is a positive integer. 

To achieve the third goal of computational difficulty, we will start with the challenge of how to choose $e$ and $d$. Let's first manipulate the equation a little bit.
Given that $ed = k\phi(n)+1$, when the modulus is $\phi(n)$, we have 
\begin{equation}
ed \equiv k\phi(n)+1 \equiv 1 \hspace{0.1cm} (\text{mod} \hspace{0.1cm} \phi(n)),
\end{equation} 
where the last congruent relation comes from the fact that $k$ is a positive integer. The congruence we get from the above manipulation $ed \equiv 1 \hspace{0.1cm} (\text{mod} \hspace{0.1cm} \phi(n))$ reveals that if $e$ and $\phi(n)$ are relatively prime, then $d$ is an inverse of $e$ modulo $\phi(n)$ and the existence of $d$ is guaranteed according to the Bézout's theorem (Section \ref{sec:modular}). 

Now we just need to find a number $e$ that is relatively prime to $\phi(n)$, and the corresponding inverse modulo $n$, denoted by $d$. Finding a number $e$ that is relatively prime to $\phi(n)$ should not be a difficult problem if given $\phi(n)$. Finding the corresponding inverse $d$ of $e$ modulo $\phi(n)$ could be done through \textit{the Extended Euclidean Algorithm} efficiently as $gcd(e,\phi(n)) = 1$. 

We have successfully found a way to find an appropriate $e$ and $d$. However, this does not conclude the problem. In the third goal of public-key cryptosystems, it requires that there exists no known efficient way to calculate $d$ given the information of $e$ and $n$. Obviously, we still have not reached that goal. If $n$ is not chosen carefully, an attacker might be able to easily figure out the value of $\phi(n)$ and further efficiently figure out $d$ based on $e$. 

Achieving the last goal of the public-key cryptosystems is one of the most elegant parts of the RSA algorithm. We know that there exist no known efficient method to perform prime factorisation(Section \ref{sec:factor}). If the receiver can first find two large random prime numbers $p$ and $q$ privately and let $n = p \cdot q$, then there will exist no efficient way to reverse this process to get $p$ and $q$ from only $n$. Further, it will be computationally difficult to get the value of $\phi(n)$ as stated in Section \ref{sec:euler}. 

However, it will be super easy for the valid receiver to calculate $\phi(n)$ as $\phi(n) = (p-1)(q-1)$. This is also known as the ``trap-door one-way function'', which is similar with how our shipping drop box works. 

Finally we have achieved all the three goals mentioned at the beginning. The receiver just needs to first choose two large enough prime numbers $p$ and $q$, and get $n = p \cdot q$ and $\phi(n)=(p-1)(q-1)$.  Then $p$ and $q$ can be destroyed to prevent potential leaks. The receiver can further get the public key $(e,n)$ by choosing a large enough $e$ that is relative prime to $\phi(n)$ and then the private key $(d,n)$ could be computed based on $ed \equiv1 ( mod \hspace{0.1cm} \phi(n))$. As there’s no efficient way to compute $\phi(n)$ based on $n$ as it requires a prime factorization, thus the third goal of computation difficulty will be achieved. 

We still have one last question left unanswered from Section \ref{sec:goal}. How can we ensure $n$ and $M$ to be relatively prime? Unfortunately, we cannot ensure it directly. However, we know that $n = p \cdot q$ with $p, q$ being prime, which means $n$ will be relatively prime to all numbers less than $n$ except $p, q$ and their multiples.  The only case in which $M$ and $n$ are not relatively prime is when $M$ is a multiple of $p$ or $q$ or both, which has an extremely low chance in terms of probability considering we also require $M < n$ in Goal 2. 

Up until this point, all the requirements to achieve the three goals of public-key cryptosystems are satisfied. In the following section we provide a toy example to sort out the process. 

\section{A Toy Example}\label{sec:example}

The detailed implementation specifications of the RSA algorithm in real world can be found in \cite{moriarty2016pkcs}. Suppose Alice wants to send a message ``Tue 7PM'' to Bob secretly using the RSA algorithm. First, Bob needs to decide his private key $(d,n)$ and public key $(e,n)$ for the communication. Bob will choose two large random prime numbers $p$ and $q$. Let's assume $p = 1721$ and $q = 1801$. In real world, these two numbers should be much larger such that it is unrealistic for modern computers to obtain the prime factors $p$ and $q$ from $n$.  $n$ can be computed as $n = p \cdot q = 3099521$. We can also obtain Euler's totient function of $n$ as $\phi(n) = (p-1)(q-1) = 3096000$. 

The next step for Bob is to choose a public key $e$, which is a number relatively prime to $\phi(n)$. For example, the standard sizes for RSA keys starts from 512 bits. To get a very high-strength key, the key size requires 4096 bits.  Here in our toy example we choose $e = 1012333$.  Now Bob needs to compute the private key $d$. Based on the equation $ed \equiv 1 \hspace{0.1cm} (\text{mod} \hspace{0.1cm} \phi(n))$, we could get the inverse of $e$ modulo $\phi(n)$ as $d = 997$ using the Extended Euclidean Algorithm. After $e$ and $d$ are determined, $p$ and $q$ can be destroyed or hidden for the sake of security. Bob can release his public key $(e,n)$ to the public while keep $d$ private. 

From Alice's perspective, Alice needs to first obtain Bob's public key $(e,n)$, then she could convert the message she wants to send into its numerical representations. Here we use ASCII (American Standard Code for Information Interchange) to convert ``Tue 7PM'' into numerical representation as: 084 117 101 032 055 080 077. 

If the message is too long, Alice could divide the message into smaller blocks, then encode each block separately. Here we divide the message into blocks that has 3 digits in each of them. There are seven blocks in the message including the space. With the public key $(e,n) = (1012333, 3099521)$, Alice could obtain the ciphertext through $M^e \hspace{0.1cm}(\text{mod } n)$ to get $084^{1012333}  \equiv 469428 (\text{mod } 3099521), 117^{1012333}  \equiv 547387 (\text{mod } 3099521), \dots \dots.$ The complete ciphertext $C$ is shown as "0469428 0547387 2687822 1878793 0330764 1501041 1232817". When Bob receives the ciphertext, he will decrypt the ciphertext using his own private key $(d,n) = (997, 3099521)$ to get $0469428^{997} \equiv 84 (\text{mod } 3099521), 0547387^{997} \equiv 117(\text{mod } 3099521), \dots \dots, 1232817^{997} \equiv 77 \\(\text{mod } 3099521)$. 
Finally he recovers the original message by looking up the ASCII table to get the plaintext message ``Tue 7PM''.

\section{Student Learning Outcome Assessment}
To study the effectiveness of the proposed student-oriented approach in explaining the RSA algorithm, we conducted a comparative analysis with the traditional method outlined in \cite{rosen2019discrete}. In the traditional method, the encryption and decryption process are presented upfront to the students, followed by the corresponding proof utilising number theory knowledge to enhance comprehension of the algorithm. The explanatory style from \cite{rosen2019discrete} presents the conventional approach to teaching the RSA algorithm. 

The comparison involved two sections of the same course, namely \textit{CSC 140 Discrete Structures} at Rider University. These sections comprised 24 and 26 undergraduate students, respectively, all majoring in computer science or cybersecurity. Given that this is a 100-level course and a prerequisite for several higher-level courses, the majority of students are either freshmen or sophomores, aligning with the target readership of this paper.

In these two sections, all course content, excluding the RSA algorithm section, followed the same instructional format. Equal lecture time was allocated to each topic in both sections. Student performance was compared based on related assignment grades. Both sections were presented with identical assignment problems and grading criteria.

\begin{figure*}
\centering
\begin{subfigure}[t]{.48\textwidth} 
  \centering
\includegraphics[trim={0cm 3cm 0cm 0cm},clip,width=5cm]{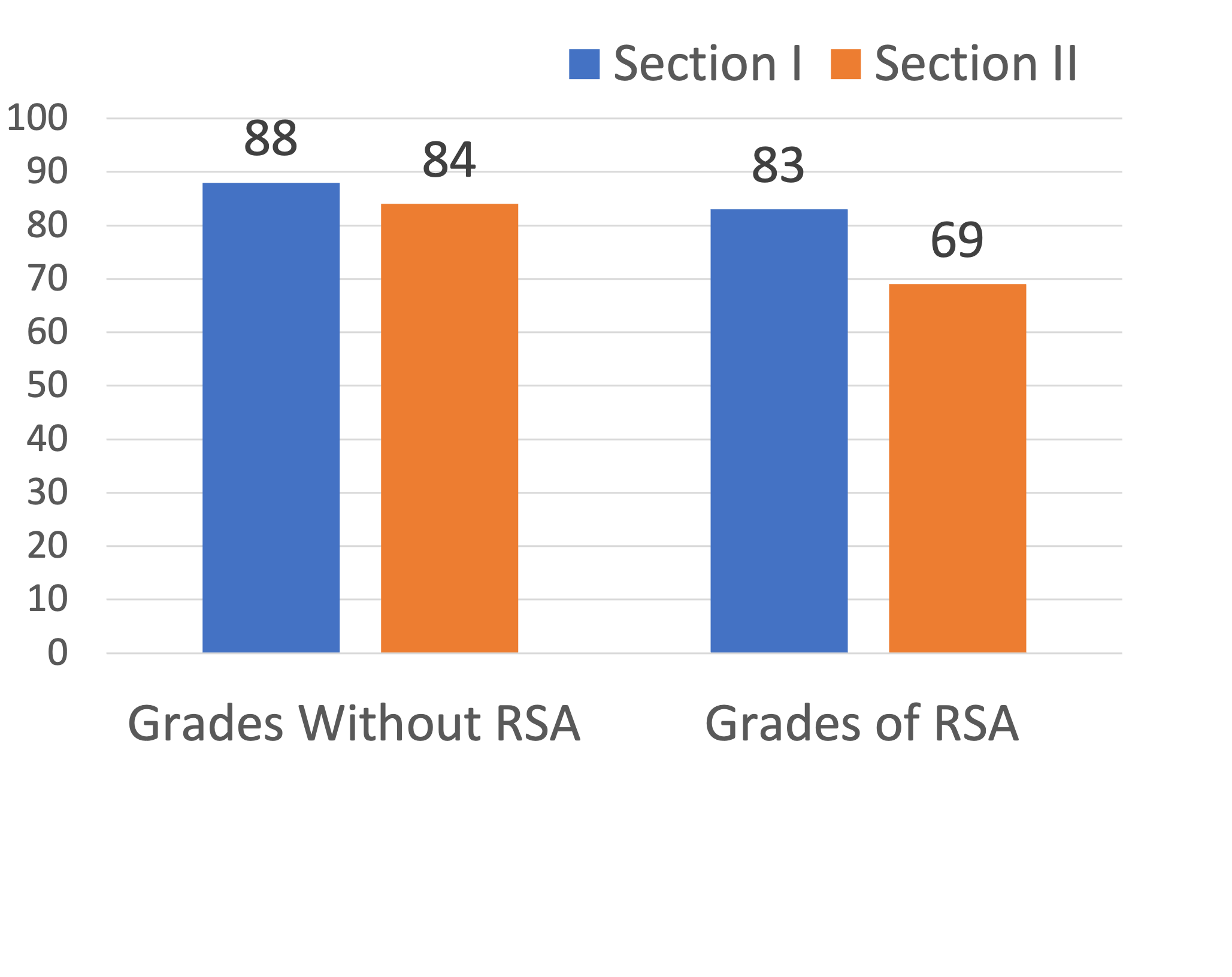}
\label{fig:result_a}
  \caption{"Grades Without RSA" refers to the average grades of assignments unrelated to the RSA algorithm, which are taught in the same manner; "Grades of RSA" represents the average grades related to the RSA algorithm, which are taught differently.}
\end{subfigure} \hfill
\begin{subfigure}[t]{.48\textwidth}
  \centering
\includegraphics[trim={0cm 0cm 0cm 0cm},clip,width=5cm]{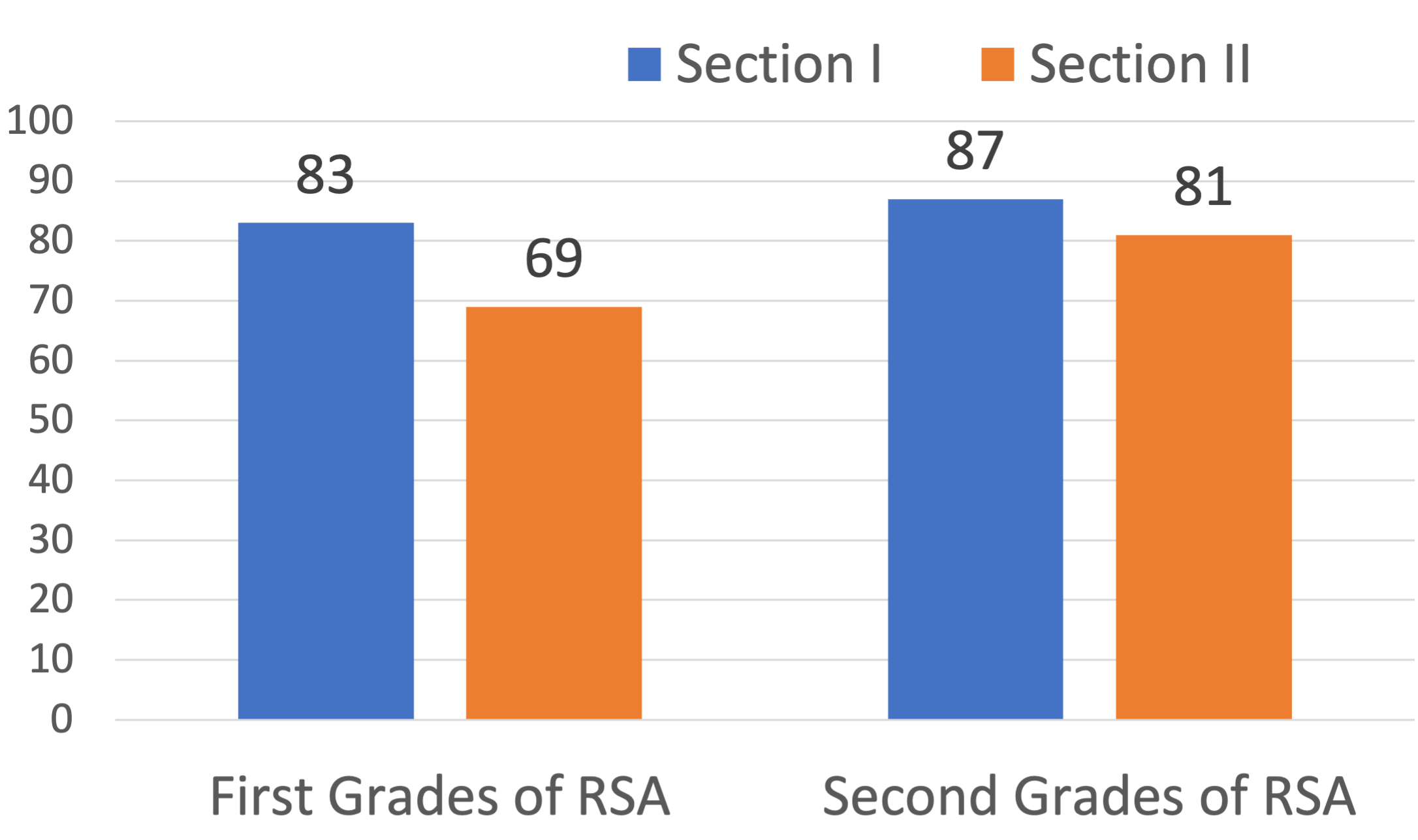}
  \caption{"First Grades of RSA" represent the averaged grades of the assignment related to the RSA algorithm for the two sections; "Second Grades of RSA" refer to the averaged grades students received after the alternative way is offered.}
\end{subfigure}
\caption{Students learning outcome comparison in terms of assignment grades from two sections of the same course.}
\label{fig:performance}
\end{figure*}

 The study involved initially employing the proposed student-oriented method outlined in this work for students in Section I and the traditional method from \cite{rosen2019discrete} for students in Section II. Subsequently, a related assignment was administered. Following this, both sections were exposed to an alternative introduction method—Section I students were presented with the traditional explanation, while Section II students were introduced to the proposed student-oriented approach. Finally, a makeup opportunity for the assignment was extended to all students. Detailed results are presented in Fig. \ref{fig:performance}.

In Fig. \ref{fig:performance} (a), we initially compared two categories of student grades: "Grades Without RSA" and "Grades of RSA." The former represents the averaged grades for all assignments throughout the semester, excluding the one related to the RSA algorithm. With a total of 9 assignments for the entire semester, all topics pertaining to these assignments are taught in the same way. Our analysis revealed that students from Section I performed, on average, 4 points higher than those from Section II (each assignment is out of 100 points). 

On the other hand, "Grades of RSA" focuses solely on the assignment related to the RSA algorithm, considering a single assignment. Our findings indicated that students in Section I outperformed those in Section II by an impressive average margin of 14 points. If the effectiveness of the teaching methods were equal for both sections, we would anticipate a much smaller average grade difference than the observed 14 points. Consequently, these results underscore the effectiveness of the student-oriented approach in explaining the RSA algorithm compared to the traditional method.

Upon offering both sections the alternative teaching method, we observed an improvement in grades for both groups (Fig. \ref{fig:performance} (b)). However, the gap in grades between the two sections narrowed from 14 points to 6 points. This reduction further validates the efficacy of the student-oriented teaching approach.

\section{Conclusion}\label{sec:conclusion}
As the significance of cybersecurity continues to rapidly increase across various facets of society, comprehending the fundamental logic behind widely used security mechanisms becomes essential not only for cybersecurity students but also for a broader audience. In this study, we present a self-contained and student-oriented interpretation of the RSA algorithm, a cornerstone in public-key cryptosystems. Beginning with three goals of public-key cryptosystems, we guide readers through a step-by-step explanation of how the RSA algorithm satisfies and implements each of these three goals. Our student learning outcome assessment, conducted across two different course sections, demonstrated the effectiveness of our approach, with an average grade difference of 14 points compared to the traditional method of teaching the RSA algorithm.We envision this work serving as a more approachable channel for readers to grasp the intricacies of the RSA algorithm. 

\small{
\printbibliography

@article{rivest1978method,
  title={A method for obtaining digital signatures and public-key cryptosystems},
  author={Rivest, Ronald L and Shamir, Adi and Adleman, Leonard},
  journal={Communications of the ACM},
  volume={21},
  number={2},
  pages={120--126},
  year={1978},
  publisher={ACM New York, NY, USA}
}

@incollection{diffie2022new,
  title={New directions in cryptography},
  author={Diffie, Whitfield and Hellman, Martin E},
  booktitle={Democratizing Cryptography: The Work of Whitfield Diffie and Martin Hellman},
  pages={365--390},
  year={2022}
}

@article{bernstein2017post,
  title={Post-quantum cryptography},
  author={Bernstein, Daniel J and Lange, Tanja},
  journal={Nature},
  volume={549},
  number={7671},
  pages={188--194},
  year={2017},
  publisher={Nature Publishing Group UK London}
}

@techreport{moriarty2016pkcs,
  title={PKCS\# 1: RSA cryptography specifications version 2.2},
  author={Moriarty, Kathleen and Kaliski, Burt and Jonsson, Jakob and Rusch, Andreas},
  year={2016}
}

@misc{katz2020introduction,
  title={Introduction to Modern Cryptography CRC Press},
  author={Katz, Jonathan and Lindell, Yehuda},
  year={2020},
  publisher={Taylor \& Francis), Boca Raton, FL, USA}
}

@book{rosen2019discrete,
  title={Discrete mathematics and its applications},
  author={Rosen, Kenneth H},
  year={2019},
  publisher={The McGraw Hill Companies,}
}

@inproceedings{shor1994algorithms,
  title={Algorithms for quantum computation: discrete logarithms and factoring},
  author={Shor, Peter W},
  booktitle={Proceedings 35th annual symposium on foundations of computer science},
  pages={124--134},
  year={1994},
  organization={Ieee}
}

@book{rosen2011elementary,
  title={Elementary number theory},
  author={Rosen, Kenneth H},
  year={2011},
  publisher={Pearson Education London}
}

@article{wahab2021hiding,
  title={Hiding data using efficient combination of RSA cryptography, and compression steganography techniques},
  author={Wahab, Osama Fouad Abdel and Khalaf, Ashraf AM and Hussein, Aziza I and Hamed, Hesham FA},
  journal={IEEE access},
  volume={9},
  pages={31805--31815},
  year={2021},
  publisher={IEEE}
}

@article{fotohi2020securing,
  title={Securing wireless sensor networks against denial-of-sleep attacks using RSA cryptography algorithm and interlock protocol},
  author={Fotohi, Reza and Firoozi Bari, Somayyeh and Yusefi, Mehdi},
  journal={International Journal of Communication Systems},
  volume={33},
  number={4},
  pages={e4234},
  year={2020},
  publisher={Wiley Online Library}
}

@article{imam2021systematic,
  title={Systematic and critical review of rsa based public key cryptographic schemes: Past and present status},
  author={Imam, Raza and Areeb, Qazi Mohammad and Alturki, Abdulrahman and Anwer, Faisal},
  journal={IEEE Access},
  volume={9},
  pages={155949--155976},
  year={2021},
  publisher={IEEE}
}

@inproceedings{liestyowati2020public,
  title={Public key cryptography},
  author={Liestyowati, Dwi},
  booktitle={Journal of Physics: Conference Series},
  volume={1477},
  number={5},
  pages={052062},
  year={2020},
  organization={IOP Publishing}
}

@inproceedings{anusha2020symmetric,
  title={Symmetric Key Algorithm in Computer security: A Review},
  author={Anusha, R and Kumar, MJ Dileep and Shetty, Vaishnavi S and Hegde, N Prajwal},
  booktitle={2020 4th International Conference on Electronics, Communication and Aerospace Technology (ICECA)},
  pages={765--769},
  year={2020},
  organization={IEEE}
}

@article{national2019quantum,
  title={Quantum computing: progress and prospects},
  author={National Academies of Sciences, Engineering, and Medicine and others},
  year={2019},
  publisher={National Academies Press}
}

@book{hidary2019quantum,
  title={Quantum computing: an applied approach},
  author={Hidary, Jack D and Hidary, Jack D},
  volume={1},
  year={2019},
  publisher={Springer}
}

@incollection{easttom2022quantum,
  title={Quantum computing and cryptography},
  author={Easttom, Chuck},
  booktitle={Modern Cryptography: Applied Mathematics for Encryption and Information Security},
  pages={397--407},
  year={2022},
  publisher={Springer}
}
}

\end{document}